\documentstyle[aps, graphicx]{revtex}

\begin{document}

\def\gv{GeV/c$^2$}
\def\cs2{CS$_2$}
\def\h2{H$_2$}
\def\40K{$^{40}$K}
\def\68Ge{$^{68}$Ge}
\def\14C{$^{14}$C}
\def\bqcm{Bq/cm$^3$}
\def\cm2{cm$^2$}
\def\cm3{cm$^3$}
\def\tenmtwo{10$^{-2}$}
\def\tenmthree{10$^{-3}$}
\def\tenmfour{10$^{-4}$}
\def\abt{$\sim$}
\def\perkgday{(kg$\cdot$day)$^{-1}$}
 
\draft
\title{Low Pressure Negative Ion Drift Chamber for Dark Matter Search}
\author{D.P. Snowden-Ifft}
\address{Physics Department, Occidental College, Los Angeles, CA
90041}
\author{C.J. Martoff}
\address{Department of Physics, Temple University, Philadelphia, PA
19122}
\author{J.M. Burwell}
\address{Physics Department, Mount Holyoke College, South Hadley, MA
01075}
\date{\today}
\maketitle
\begin{abstract}
Weakly Interacting Massive Particles (WIMPs) are an attractive 
candidate for the dark matter thought to make up the bulk of the mass 
of our universe.  We explore here the possibility of using a low 
pressure negative ion drift chamber to search for WIMPs.  The 
innovation of drifting ions, instead of electrons, allows the design 
of a detector with exceptional sensitivity to, background rejection 
from, and signature of WIMPs.
\end{abstract}


Since the earliest astrophysical measurements on galaxy
clusters during the 1930's an observational problem
of dark mass has existed\cite{Zwicky}.  This problem persists to 
the present day in the most recent space based measurements on galaxy
clusters\cite{xraygas}. In fact, the problem is present in 
practically every structure studied which is the size of a galaxy
or larger\cite{Trimble}.  Orbital velocities in large systems
are systematically larger than they should be in the gravitational
potential well of the visible mass in these systems.  The 
discrepancies are not subtle.  Even the lowest estimates imply that 
there is several times more dark gravitating material than there is
visible matter~\cite{Trimble}.  The existence of dark matter is hardly 
in doubt; the question is, ``What is it?'' 

The Big Bang theory of cosmology and the Standard Model of particle
physics provide some clues.  Big Bang nucleosynthesis
calculations indicate that most of the dark matter
must be non-baryonic~\cite{Turner:9811454}.  
When other arguments and data from
cosmology and particle physics are included a general consensus in 
the field emerges that there are now 4 strong contenders\cite{Dodelson}: 
Massive Compact Halo Objects (MACHOs), neutrinos,
axions and Weakly Interacting Massive Particles (WIMPs).  
Our concern here is with WIMP dark matter.  Primack, Seckel, and 
Sadoulet~\cite{Primack} argued that if a stable WIMP exists in nature 
it {\em must}
make up the dark matter.  Supersymmetry provides a theoretical 
home~\cite{Jungman} for WIMPs and the lightest-super-partner is 
stable in most theoretical schemes.  
 
A considerable experimental effort to detect this attractive dark
matter candidate has
been mounted in the last two decades~\cite{Jungman}.  All direct 
searches for WIMPs utilize the same principle.  WIMPs are detected by 
the recoiling ions they produce within a target material.  The 
challenge is that the recoil energies ($\sim1$ keV/amu) and interaction 
rates ($<1$ \perkgday) are both small.  A close look at the DAMA 
experiment~\cite{Bernabei:1998} is illustrative of the search for dark 
matter today.  First, the ``no counts'' limit, 5$\times$10$^{-4}$ 
\perkgday (defined as 2.3 counts divided by the exposure in kg-days) is 
several orders of magnitude smaller than the published limit of 13 \perkgday.  
In other words, DAMA is currently background limited.  
{\em In fact all currently running dark matter experiments
are limited not by exposure (mass$\times$time) but by background
levels in the detectors.  }Since background, not exposure, is the 
limiting factor, most experiments utilize some form of event 
discrimination to reduce the integral background (IBG) to 
some lower accepted background (ABG).  DAMA uses pulse shape 
discrimination to lower the ABG to $\sim \,$1/3.5 of the IBG, yielding
the above limit.  In the presence of non-zero ABG the 
only way to {\em detect} WIMPs is to look for some signature of 
WIMPs in the data.  Without the ability
to sense the direction of the recoil (extremely difficult in solids
and liquids where the range is of order 100 \AA)
the only available signature is
a small annual modulation of the total rate and energy spectrum\cite{Spergel}.  
Currently the DAMA collaboration claims to have found such a
modulation in their data~\cite{Bernabei:1998}.  
The lessons to be drawn from this discussion are as follows.  First, 
large mass detectors
are not at present an absolute requirement for improving limits because
background, not exposure, still dominates the best experiments.  
Second, background rejection factors play a large role in
current limits.  And finally the presence of non-zero ABG leads to
a desire for for a strong signature.  We are proposing to build a
detector which directly addresses this current experimental situation.  

The detector will be a low pressure TPC 
filled with a mixture of target gas and an electronegative gas.   
The use of electronegative gas permits the chamber to 
operate in a new mode creating a Negative Ion Drift Chamber (NIDC).  
In a NIDC primary ionization electrons must be
rapidly and efficiently
captured by the electronegative gas molecules.  The resulting
negative ions drift to the anode.   In the strong, inhomogeneous
field near the anode wires the ions must be
field ionized so that avalanche multiplication can occur.  This last
characteristic restricts the possible electronegative gases that
can be used.  
Single-wire proportional counters using negative ion drift were previously
used by one group~\cite{Crane} and gas detectors have been considered
for dark matter searches before\cite{Buckland}.  The innovation
here is that we realized and have verified experimentally that a NIDC has
a number of specific advantages when applied to the search for dark matter.  

The {\em raison d'etre} of a gaseous dark matter detector is its
ability to measure components of the range of a WIMP-recoil.  The
NIDC concept greatly helps in this regard.  Because ions drift,
instead of electrons, transverse {\em and} longitudinal diffusion are
suppressed to thermal levels,
$0.72 {\rm mm}\,\, \sqrt{(L/1\,{\rm m}) \times (1 {\rm kVcm^{-1}}/E)}$.  
No magnetic field is needed to suppress diffusion making
the detector easily scalable.   
Standard methods of measuring the components of the range parallel to
the anode plane can be utilized.  
The slow ion drift allows the track length projection in the drift direction
to be measured with high resolution, even for very short tracks.   

Several prototype NIDCs have been built and successfully operated 
in our labs using \cs2 as the electronegative component.  
Data have been taken with a variety of low pressure 
Ar:\cs2 and Xe:\cs2 gas mixtures.  These chambers run stably with
drift fields at least as high as 3200 V/cm at 40 Torr.  
The capture distance for ionization electrons has been measured to be
a few tenths of a millimeter at 40 Torr, and the lateral and longitudinal 
diffusion of \cs2$^{-}$ at the thermal limit has been experimentally confirmed.  
The NIDC concept works.  

The active volume of the first Directional Recoil Identification From 
Tracks (DRIFT) detector will be $\sim$1 m$^3$ in the form of two 
back-to-back TPCs sharing a common cathode within a cylindrical 
vacuum vessel.  The gain structure will be a multiwire proportional 
chamber (MWPC) with 20(100)$\mu$m anode(grid) wires spaced at 1 mm.  
MSGD devices with back-side PGA output connection\cite{Tanimori} are 
also under consideration.  The fiducial region for WIMP recoil events
is the volume between the cathode and the grid wires.  The grid will be
segmented with an annulus around the edge to veto ionizing radiation
entering from the sides.  Ionizing radiation entering from the top and bottom
has to pass through the MWPCs where it produces a characteristic fast
pulse shape (due to the high electric fields there) and can therefore
also be vetoed.  

\begin{figure}[h] 
\centerline{\includegraphics[width=3.0in,angle=0]{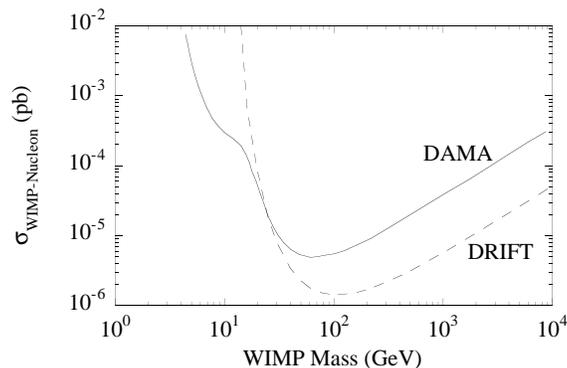}}
\caption{Upper limits (90\% c.l.) on the spin independent WIMP-nucleon 
interaction cross section obtained by the DAMA collaboration in 
comparison to the sensitivity of DRIFT after one year of running.  
Halo parameters and coherence parameterizations were identical.  The
DRIFT threshold for this calculation was 40 keV.} 
\label{fig:Limits} 
\end{figure}

To determine the sensitivity of the DRIFT detector a Monte
Carlo simulation was run in which WIMP-Ar scattering events were 
generated in pure Ar.  We do not intended to run DRIFT with a pure Ar fill.  
Ar is radioactive, it is unquenched, it is not electronegative, its
atomic number is not high enough to make it a good WIMP scattering target, 
and it has no spin.  However, the range
and ionization of low energy ions in pure Ar are known\cite{Cano}.  
The measurement of these quantities for more palatable gas mixtures
is a high priority for the DRIFT collaboration.  Until then we will
use Ar as illustrative of the DRIFT concept.  
The gas pressure must be kept low, we will use 40 Torr, in order that the
range of a typical WIMP recoil is long enough to be measured using
anodes spaced at 1 mm.  The number of Ar recoils in the 1 m$^3$ DRIFT 
detector run for one year with energy greater than 40
keV was calculated as a function of WIMP mass.   
A plot of the resulting sensitivity is shown in Figure \ref{fig:Limits}.   
The curves shown are the upper limits on the WIMP-nucleon
(spin-independent) scattering cross section that could be set if no 
nuclear recoils were detected in one year.  Note that the limit from 
running the DRIFT detector for one year with 40 Torr Ar would be roughly
$\times 5$ stronger than the current DAMA limits\cite{Bernabei:1996} 
for large mass WIMPs even though the mass of Ar in DRIFT is only 0.094 kg.  
The DRIFT limit would continue to improve as 1/t until the exposure
reached the order of 1/ABG.  This discussion shows that high sensitivity can be 
achieved with a very low target mass if the backgrounds can be 
sufficiently reduced.  

Experience has shown\cite{Smith&Lewin:1990}
that it is practically impossible to keep all background radiation 
out of the detector volume.  The only real hope for achieving the
sensitivity indicated above is to reject those events which do 
occur.  DRIFT has very strong background rejection capabilities because 
of its ability to measure 
several, if not all, components of the range event-by-event.   

\begin{figure}[h]
\hbox{
\includegraphics[width=5.5cm,angle=0]{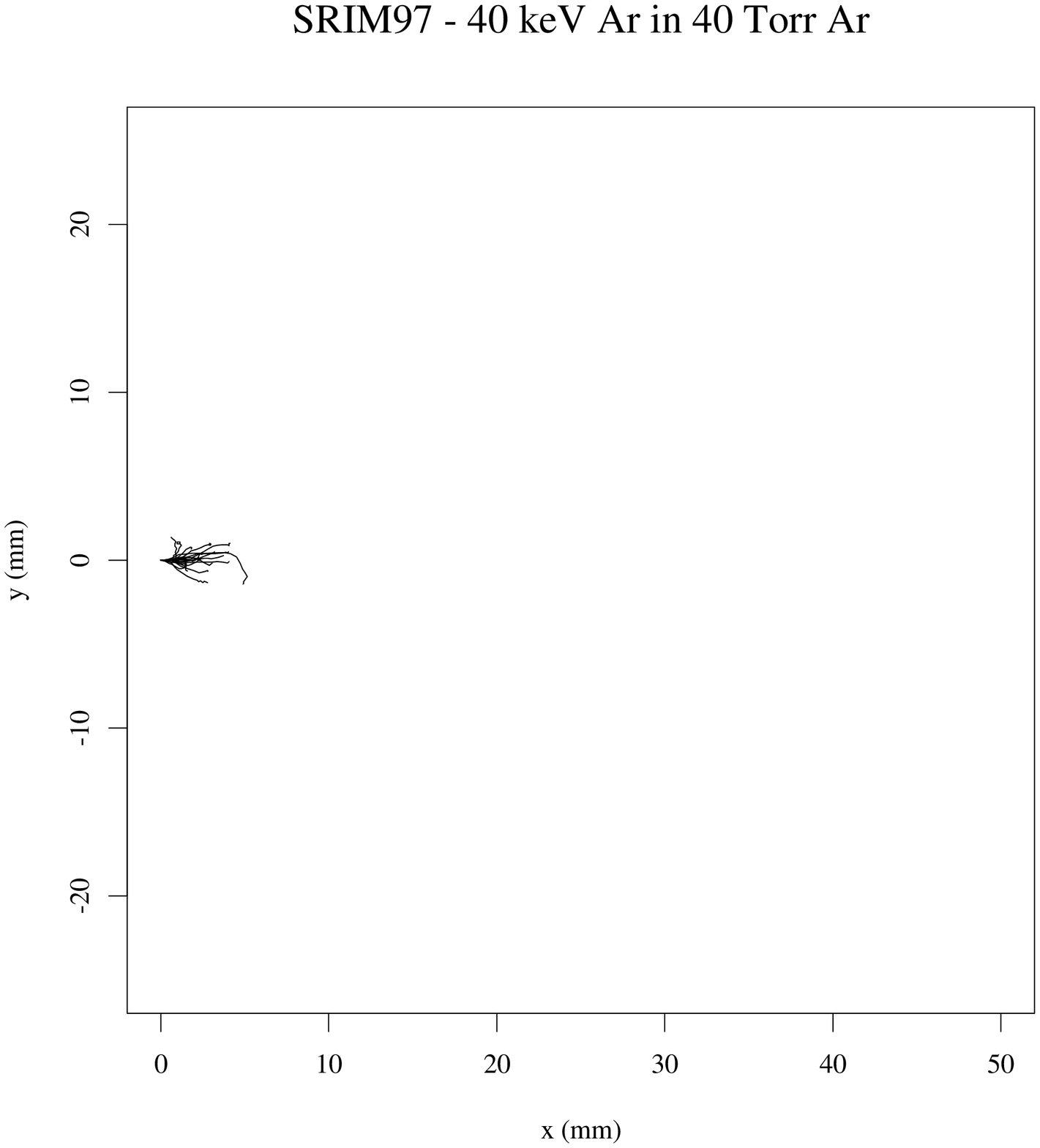}
\includegraphics[width=5.5cm,angle=0]{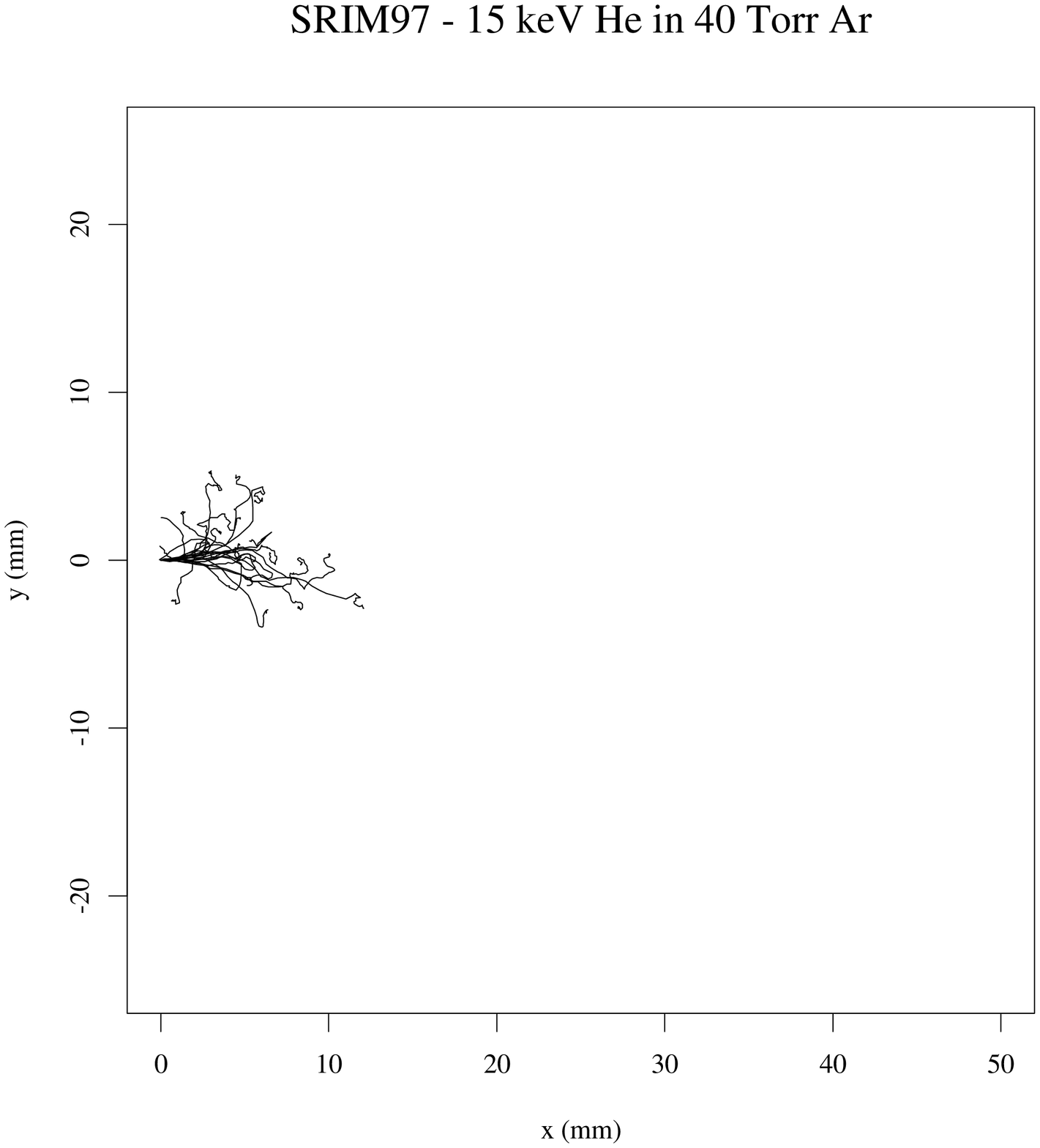}
\includegraphics[width=5.5cm,angle=0]{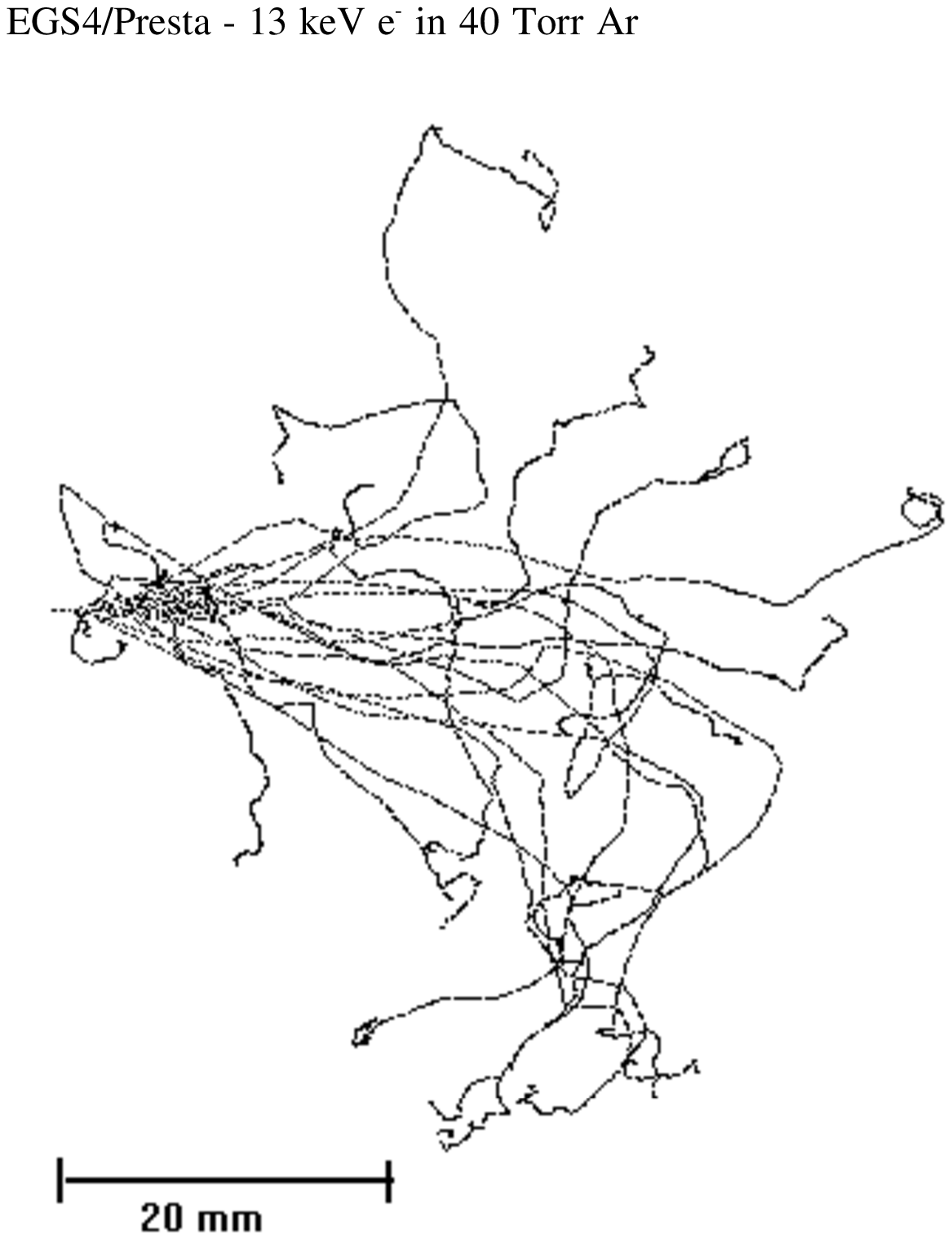}
}
\caption{
The figures above show, from left to right, 40 keV Argon recoils, 
15 keV alphas and 13 keV electrons in 40 Torr Ar.  
}
\label{fig:Backgrounds} 
\end{figure}

The $\sim$MeV alpha particles from radioactive decay of U and Th
pose a serious background problem for DRIFT.   
Alpha's which enter the fiducial volume from the 
sides will be vetoed as discussed above.  
Alphas from the grid wires or cathode which enter the fiducial
volume, however, cannot be vetoed in this way.  In addition because
the grid wires and cathode are thick relative to the range of the
alpha particles some of the alphas will emerge
with energies small enough to produce ionization equivalent to WIMP
events.  The residual range
of these particles, however, will be vastly different from
that of WIMP recoils depositing the same total ionization.  
In 40 Torr Ar a 40 keV Ar recoil produces 500 primary
ion pairs~\cite{Phipps} and has a projected range of
2.7mm~\cite{Cano}.  In contrast an alpha-particle which produces 
500 primary ionizations (15 keV~\cite{Evans}) has a projected range
of about 17mm~\cite{Evans}.  Measuring the range is a powerful
discriminant!The situation gets
complicated when one realizes that alphas do not travel in
straight lines at these energies, see Figure \ref{fig:Backgrounds}.  
The Ar recoil and alpha tracks shown in this figure were generated 
with the SRIM97 Monte Carlo~\cite{Ziegler} scaled to match
experimental data.  Cuts on just {\em two} components of the range indicate 
the an alpha mis-identification probability (MIP) less than 5\%.  
More sophisticated cuts or measurements of the third dimension will allow 
for even better alpha rejection.   
 
For the Compton events generated by photons interacting within its
fiducial volume or betas which enter the fiducial volume from the grid
wires or the cathode the DRIFT detector has an even lower MIP.   
A 13 keV electron will give 500 primary ionizations in Ar but will travel 
$\sim$85mm~\cite{Sauli_Bible}.  As with alphas,
the electrons don't travel in straight lines, see Figure \ref{fig:Backgrounds}.  
The electrons in this figure were generated using the
EGS/PRESTA~\cite{Nelson,Bielajew} simulation code.   
Cuts on just two components of the range indicate 
an electron MIP less than $3 \times 10^{-5}$.  
Again more sophisticated cuts
or measurements of the third dimension will allow for even 
better electron rejection.  Experimentally we have measured a 
mis-identification probability less than 0.001 using $^{55}$Fe (6 keV) X-rays.  

There is no rejection factor for neutrons as the recoils they 
produce are almost identical to those produced by WIMPs.  The only 
hope for seeing one or less neutron recoils per year in DRIFT is to 
insure adequate shielding.  

With the above electron and alpha MIPs in hand we can estimate the ABG in
DRIFT after one year of running.  For alphas the most important
consideration is the radiopurity of the wires and the cathode.   
Commercially available stainless steel wires have
been tested with
U and Th concentration less
than 0.5 ppb~\cite{SpoonerPC}.  Acrylic to form the central cathode can be
had with U and Th
contamination 
less than 0.01 ppb~\cite{SNO}.   
The upper limits on the the radiopurity of the these elements and the 
MIP for alphas give an upper limit 
of the order of 10 events per year from alpha background in DRIFT.  For the 
case that the actual levels are comparable to this limit, further 
highly effective reduction strategies are under study.  The most promising
is to measure the time difference between
the arrival of the \cs2$^{-}$ ions at the anode and the arrival of the
\cs2$^{+}$ ions at the cathode\cite{NygrenPC}.  This will allow
events to be localized in the gas away from the wires and cathode.  
 
Events mis-identified as WIMP recoils can also arise from low energy
Compton electrons from gammas which leak through the shield or are
generated by the shield or detector materials, or from 
beta decays in the wires and gas.  Using published
flux measurements of gammas inside operating dark matter
shields~\cite{Smith&Lewin:1990} and
measurements of beta and gamma emitters in various construction
materials we have been able to estimate the IBG from these
sources.  The $3 \times 10^{-5}$ upper limit on the MIP
for low energy electrons means that the ABG from electrons
will be less than 0.03 events per year.  
 
There is no rejection factor for neutrons
but by going underground the number of expected neutron
recoils can be reduced to levels far below 1 event per
year~\cite{Smith&Lewin:1990}.  In
summary there is every expectation that DRIFT will achieve the
``zero-background" sensitivity discussed above.  
``What if we see something consistent with a WIMP recoil?''

In addition to having good sensitivity, 
the DRIFT detector also has a very robust signature for
{\em detecting} WIMPs.  WIMP signatures in direct detection 
experiments\cite{Spergel} arise from the fact that the solar 
system rotates
around the center of the Galaxy, through a 
halo of WIMPS generally believed to be {\em nonrotating}.  
WIMP velocities
relative to the Earth are therefore a combination of the WIMP's own
Maxwellian, isotropic velocity distribution in the galactic potential 
well plus a uniform velocity 
(approximately equal in magnitude to the rms WIMP speed) due
to the Solar System's rotation
around the center of the Galaxy.  This
motion (R.A. 21hr 12.0', Dec. +48.19$^{\circ}$) is roughly toward the 
constellation 
Cygnus so colloquially one can say there is a WIMP ``wind'' blowing 
at the Earth from the direction of Cygnus.  
 
Dark matter detectors which only measure
energy deposition 
can take advantage of this asymmetric velocity 
distribution in the following way.  From April-September the Earth's 
velocity vector around the Sun has a component which is ``into the 
wind'' causing higher energy recoils at a higher rate in the
detector.   
During the months October-March the opposite occurs.  For a threshold
energy of the recoil, 
E$_{th}$ = 0 keV this asymmetry (difference divided by the sum)
is 2\% while for E$_{th} = \frac{1 keV}{amu} A$ it rises to 5\%\cite{Spergel}.  

Direction-sensitive experiments, like DRIFT,
give a much stronger signature associated with the WIMP wind
directional asymmetry.   
Consider a drift chamber located at North latitude equal to the declination of 
Cygnus.    Over a sidereal day 
the direction of the WIMP ``wind'' relative to the anode
plane changes from normal ($\tau_{cyg}=90^{\circ}$) to parallel
($\tau_{cyg}\approx0^{\circ}$).    By measuring two
components of the recoiling atom's range, $\Delta x$ and $\Delta z$ 
the projected angle $\tau_{recoil}$ 
can be calculated for each event.  
We have extended the work of Spergel\cite{Spergel} by performing a 
Monte Carlo simulation of the recoils produced from
100 GeV WIMPs in 40 Torr Ar, 
including SRIM-generated\cite{SRIM} scattering of the Ar recoils and 
drift diffusion.  
The simulation shows that 
$\Delta \tau = \mid \tau_{recoil} - \tau_{cyg} \mid$
has an asymmetric
distribution peaked near 0$^{\circ}$.   An
asymmetry is formed by counting recoils with $\Delta \tau$ less
than 45$^{\circ}$ and greater than
45$^{\circ}$.    For E$_{th}$ = 0 keV 
this asymmetry is 7\% while for 
E$_{th} = \frac{1 keV}{amu} A$ it rises to 17\%.  
Much larger 
asymmetries, approaching 1, can be had if the start of the
track can be distinguished from the end of the track.  This is
certainly possible in principle since the ionization per unit 
length does change over the length of the track.  
 
Even without ``head-tail" discrimination, this sidereal day
asymmetry provides a very
robust signature.    There are several reasons for this.  In any
experiment seeking an effect through detection of an asymmetry
the figure of merit (FOM) to achieve any given
confidence level of detection is proportional to
$FOM \propto a^{2}N_{WIMP} / (1+\frac{N_{\rm background}}{N_{\rm
WIMP}})$
where $N_{WIMP}$ is the number of WIMP events detected 
and $N_{background}$ is the number of background events detected.  
Clearly large mass detectors run for
long times are desirable since
this increases $N_{WIMP}$.  This is the
approach taken by groups using very large detector masses and 
seeking annual modulation.  But this expression
also shows that large asymmetries ($a^{2}$ dependence) and low
backgrounds contribute strongly to the FOM.  
As an illustration of how powerful these considerations can be
consider the DAMA report~\cite{Bernabei:1998}.  DAMA reports a
signal consistent, at the 90\% confidence level, with 
detection of an annual modulation.  In the energy range 2--12 keV, 
the modulation reported is 0.037$\pm$0.008 counts/day/kg/keV.  
This implies $\approx$ 85,000 WIMP events are in the data sample.  
A DRIFT detector with an asymmetry of 17\%,
instead of 2\%, and zero background could reach the same FOM
level with only 70 WIMP events, i.e. with $\approx$ 1000 times less exposure.   
In addition a sidereal-rate modulation rapidly goes out 
of synch with the solar day/night cycle and the short period imposes 
less stringent requirements on long term stability
of the experiment.   

For all of the reasons discussed above we feel that DRIFT is 
a powerful detector capable of making a significant contribution to
cosmology and particle physics.  

The authors would like to thank N. Spooner, P. Smith, M. Lehner and 
the other members of the UKDMC for their help during the development 
of the DRIFT concept.  
This work was supported by the Research Corporation under grant 
CC4512, the National Science Foundation under grant NSF-SGER 9808270, 
Occidental College and Temple University.  

\bibliography{Dans}

\begin{thebibliography}{10}

\bibitem{Zwicky}
F. Zwicky, Helv. Phys. Acta {\bf 6},  110  (1933).

\bibitem{xraygas}
A.~E. Evrard {\it et~al.}, Mon. Not. Roy. Ast. Soc. {\bf 292},  289  (1997).

\bibitem{Trimble}
V. Trimble, Ann. Rev. Astron. Astrophys. {\bf 25},  425  (1987).

\bibitem{Turner:9811454}
M.~S. Turner, astro-ph/9811454 (unpublished).

\bibitem{Dodelson}
S. Dodelson, E.~I. Gates, and M.~S. Turner, Science {\bf 276},  69  (1996).

\bibitem{Primack}
J.~R. Primack, D. Seckel, and B. Sadoulet, Ann. Rev. Nucl. Part. Sci. {\bf 38},
   251  (1988).

\bibitem{Jungman}
G. Jungman, M. Kamionkowski, and K. Griest, Physics Reports {\bf 267},  195
  (1996).

\bibitem{Bernabei:1998}
R. Bernabei {\it et~al.}, Phys. Lett. B {\bf 424},  195  (1998).

\bibitem{Spergel}
D.~N. Spergel, Phys. Rev. D {\bf 37},  1353  (1988).

\bibitem{Crane}
H.~R. Crane, Rev. Sci. Inst. {\bf 32},  953  (1961).

\bibitem{Buckland}
K. Buckland, M.~J. Lehner, G.~E. Masek, and M. Mojaver, Phys. Rev. Lett. {\bf
  73},  1067  (1994).

\bibitem{Tanimori}
T. Tanimori, A. Ochi, S. Minami, and T. Nagae, Nucl. Inst. Meth. A {\bf 381},
  280  (1996).

\bibitem{Cano}
G.~L. Cano, Physical Review {\bf 169},  277  (1968).

\bibitem{Bernabei:1996}
R. Bernabei {\it et~al.}, Phys. Lett. B {\bf 389},  757  (1996).

\bibitem{Smith&Lewin:1990}
P.~F. Smith and J.~D. Lewin, Phys. Rep. {\bf 187},  203  (1990).

\bibitem{Phipps}
J.~A. Phipps, J.~W. Boring, and R.~A. Lowry, Phys. Rev. A {\bf 135},  36
  (1964).

\bibitem{Evans}
G.~E. Evans, P.~M. Stier, and C.~F. Barnett, Phys. Rev. Lett. {\bf 90},  825
  (1953).

\bibitem{Ziegler}
J. Ziegler, J. Biersack, and U. Littmark, {\em The Stopping and Range of Ions
  in Solids}, 1st  ed. (Pergamon Press, Oxford, England, 1985).

\bibitem{Sauli_Bible}
F. Sauli,  in {\em Experimental Techniques in High Energy Physics}, edited by
  T. Ferbel (Addison-Wesley Publishing, ADDRESS, 1987), p.\ 81.

\bibitem{Nelson}
W. Nelson, H. Hirayama, and D. Rogers, Technical report, {SLAC},
  (unpublished), report No. 265.

\bibitem{Bielajew}
A. Bielajew and D. Rogers, NIM B {\bf 18},  165  (1987).

\bibitem{SpoonerPC}
N. Spooner, 1998, priv. comm. to PIs.

\bibitem{SNO}
G.~T. Evans {\it et~al.}, Technical report, Sudbury Neutrino Observatory,
  (unpublished), sNO-87-12, pp. 73 ff.

\bibitem{NygrenPC}
D. Nygren, 1998, priv. comm. to PIs.

\bibitem{SRIM}
J.~F. Ziegler, computer program distributed via Ziegler@Watson.IBM.com,
  IBM-Research, Yorktown, N.Y. (unpublished).

\end{thebibliography}

\bibliographystyle{prsty} 

\end{document}